\pdfoutput=1
\documentclass[prl,twocolumn,showpacs,amsmath,amssymb,superscriptaddress,floatfix]{revtex4}

\usepackage{graphicx}
\usepackage{dcolumn}
\usepackage{bm}
\usepackage{epsfig}

\begin{document}

\title{DNA nucleotide-specific modulation of $\mu$A transverse edge currents through a metallic graphene nanoribbon with a nanopore}

\author{Kamal K. Saha}
\affiliation{Department of Physics and Astronomy, University of Delaware, Newark, DE 19716-2570, USA}
\author{Marija Drndi\'{c}}
\affiliation{Department of Physics and Astronomy, University of Pennsylvania, Philadelphia, PA 19104, USA}
\author{Branislav K. Nikoli\' c}
\email{bnikolic@udel.edu}
\affiliation{Department of Physics and Astronomy, University of Delaware, Newark, DE 19716-2570, USA}

\begin{abstract}
We propose two-terminal devices for DNA sequencing which consist of a metallic graphene nanoribbon with zigzag edges (ZGNR) and a  nanopore in its interior through which the DNA molecule is translocated. Using the nonequilibrium  Green functions combined with  density functional theory, we demonstrate  that  each of the four DNA nucleotides inserted into  the nanopore, whose edge carbon atoms are passivated by either hydrogen or nitrogen, will  lead to a unique change in the device conductance.  Unlike other recent biosensors based on transverse electronic transport through DNA nucleotides, which utilize small  (of the order of pA) tunneling current across a nanogap or a nanopore yielding a poor signal-to-noise ratio, our device concept relies on the fact that in ZGNRs local current density is peaked around the edges so that drilling a nanopore away from the edges will not diminish the conductance. Inserting a DNA nucleotide into the nanopore affects the charge density in the surrounding area, thereby modulating edge conduction currents  whose magnitude is of the order of $\mu$A at bias voltage  \mbox{$\simeq 0.1$ V}. The proposed biosensor is not limited to ZGNRs and it could be realized with other nanowires supporting transverse edge currents, such as chiral GNRs or wires made of two-dimensional topological insulators.
\end{abstract}

\maketitle

The successful realization of fast and low-cost methods for reading the sequence of nucleotides in DNA is envisaged to lead to personalized medicine and applications in various subfields of genetics. The solid-state nanopores represent one of the pillars of the so-called third generation sequencing~\cite{Schadt2010}. The key issues in this approach revolve around  how to slow down the translocation speed of DNA and how to achieve single-nucleotide resolution.

The very recent experiments~\cite{Merchant2010,Schneider2010,Garaj2010} on DNA translocation through graphene nanopores have introduced a new contender into this arena. Graphene---the recently discovered~\cite{Geim2009} two-dimensional allotrope of carbon whose atoms are densely packed into a honeycomb lattice---brings its unique electronic and mechanical properties into the search for an optimal nanoelectronic biosensor. Since single layer graphene is only one-atom-thick, the entire thickness of the nanopore through which DNA is threaded is comparable to the dimensions of DNA nucleotides. Therefore,  there is only one recognition point rather than multiple contacts with DNA in the nanopore.

\begin{figure}
\includegraphics[scale=0.35,angle=0]{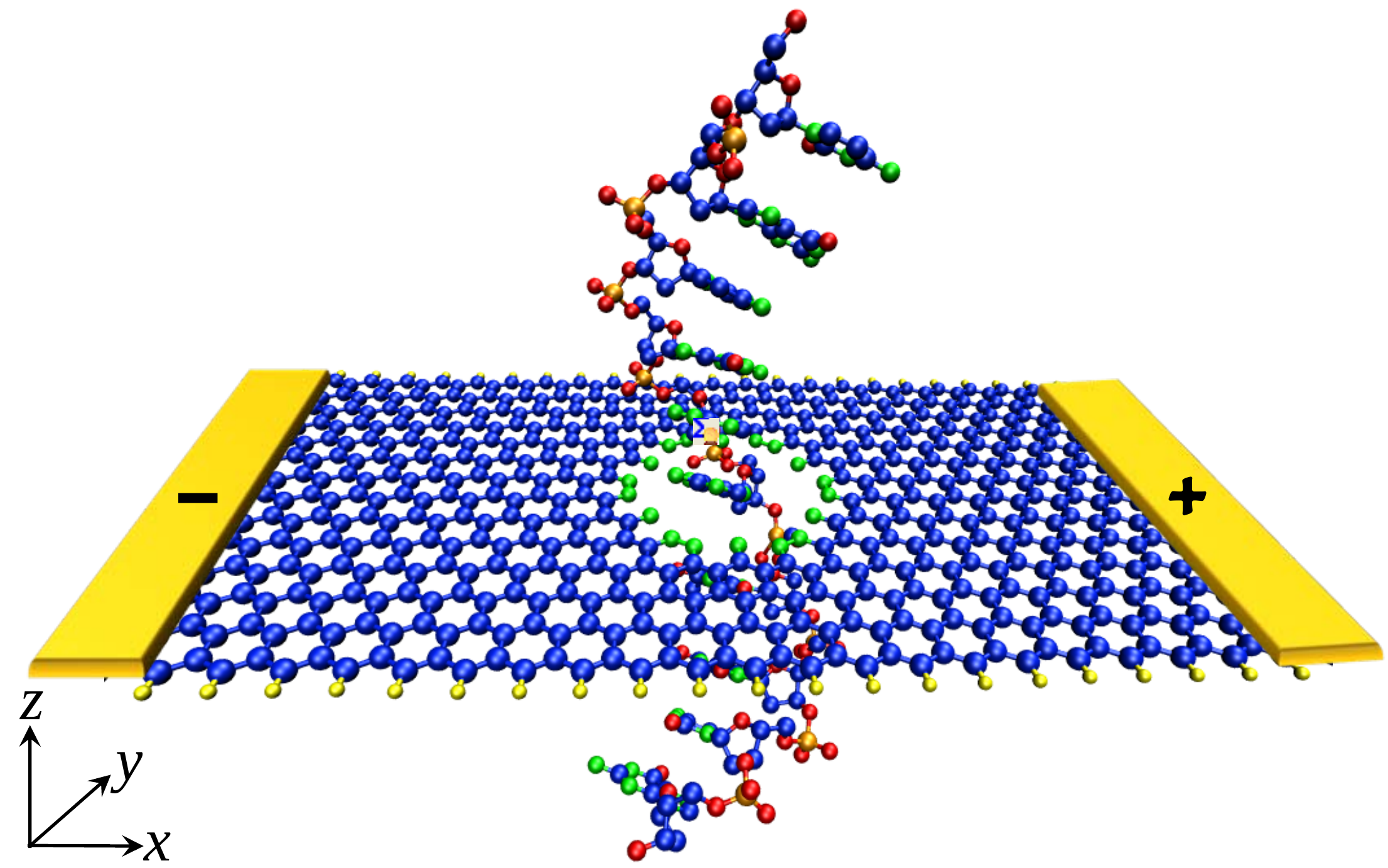}
\caption{Schematic view of the proposed two-terminal device where transverse conduction current flows around the zigzag edges of a metallic graphene nanoribbon with a nanopore, while DNA molecule is translocated through the pore to induce nucleotide-specific-modulation of such edge currents. The device central region, which is simulated via NEGF-DFT formalism, consists of 14-ZGNR (composed of 14-zigzag chains which determine its width $\approx$ 3.1 nm) and a nanopore of $\approx 1.2$ nm diameter. The edge carbon atoms of the nanopore are passivated by either hydrogen or nitrogen, while edge atoms of ZGNR itself are passivated by hydrogen. The total number of simulated atoms (C-blue, H-yellow, N-green, O-red, P-orange) in the central region, including the nucleotide within the nanopore, is around 700.}
\label{fig:setup}
\end{figure}

However, the recent experiments~\cite{Merchant2010,Schneider2010,Garaj2010} on nanopores within single or multilayer large-area graphene, which have measured fluctuations in the vertical ionic current flow due to DNA translocation through the pore, have not reached sufficient resolution to detect and identify individual nucleotides. An alternative scheme is to adapt the transverse current approach to graphene-based biosensors~\cite{Postma2010,Prasongkit2011,He2011,Nelson2010}. The past several years have seen a number of theoretical proposals~\cite{Zwolak2008,Meunier2008} and experiments~\cite{Tsutsui2010,Huang2010} on nanogaps between two metallic electrodes where the longitudinally translocated DNA through the gap modulates the transverse {\em tunneling} current. Also, the recent first-principles simulations have analyzed such modulation of the tunneling current for a nanogap~\cite{Prasongkit2011,He2011}  between metallic GNRs with zigzag edges (ZGNR) or a nanopore~\cite{Nelson2010} within semiconducting graphene nanoribbons with armchair edges (AGNR).

However, the tunneling-current based graphene biosensors will face the same challenges~\cite{Meunier2008} encountered by current experimental efforts to utilize transverse tunneling current across a gap between two gold electrodes~\cite{Tsutsui2010,Huang2010}, such as poor signal-to-noise ratio at small bias voltages due to the fact that molecular eigenlevels are typically far away from the Fermi energy of the electrodes. In this case, the tunneling is off-resonant and currents are of the order of pA at typically applied bias voltage \mbox{$\simeq 0.5$ V}~\cite{Tsutsui2010,Huang2010}. Such small off-resonant tunneling currents are highly dependent on difficult-to-control relative geometry between the molecule and electrodes, so that recent experiments have measured broad current distributions corresponding to each nucleotide in the case of bare gold electrodes~\cite{Tsutsui2010} and somewhat narrower but still overlapping distributions~\cite{Huang2010} for functionalized gold electrodes. Similarly, first-principles simulations of tunneling through the nanogap hosting a DNA nucleotide between two metallic GNRs have revealed current variation over several orders of magnitude (e.g.,  \mbox{ $10^{-2}$--$10^{-10}$ nA} at bias voltage 1 V~\cite{Prasongkit2011}) when changing the position and orientation of  nucleotides within the gap.

\begin{figure}
\includegraphics[scale=0.3,angle=0]{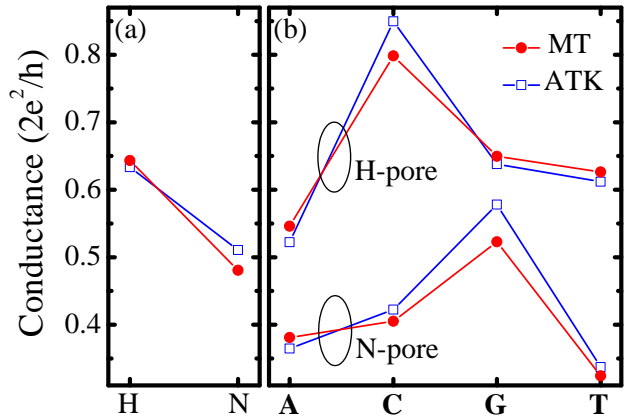}
\caption{(a) The room-temperature conductance of the two-terminal 14-ZGNRs with $\approx 1.2$ nm diameter nanopore whose edge carbon atoms are passivated by either hydrogen (H-pore) or nitrogen (N-pore). (b) The room-temperature conductance of the same device as in panel (a) when one of the four DNA nucleotides (A-adenine, C-cytosine, G-guanine, T-thymine) is inserted into the center of the nanopore within the $yz$-plane [Fig.~\ref{fig:fig5}(e)]. These conductances are computed via first-principles quantum transport simulations where both panels compare results obtained using two different NEGF-DFT codes---our home-grown MT-NEGF-DFT~\cite{Saha2010,Saha2009a} and commercial ATK~\cite{quantumwise}.}
\label{fig:fig2}
\end{figure}

The theoretical proposals to increase  transverse current across the nanogap, as in the case of carbon nanotube electrodes terminated with nitrogen where introduction of states  closer to the Fermi level enables quasiresonant tunneling, offer  only slight improvement---nA current at \mbox{$\simeq 0.4$ V} bias voltage~\cite{Meunier2008}. Applying higher bias voltage to increase the current signal is detrimental since it can lead to attraction of the negatively charged DNA backbone toward one of the electrodes thereby impeding the translocation.

\begin{figure}
\includegraphics[scale=0.3,angle=0]{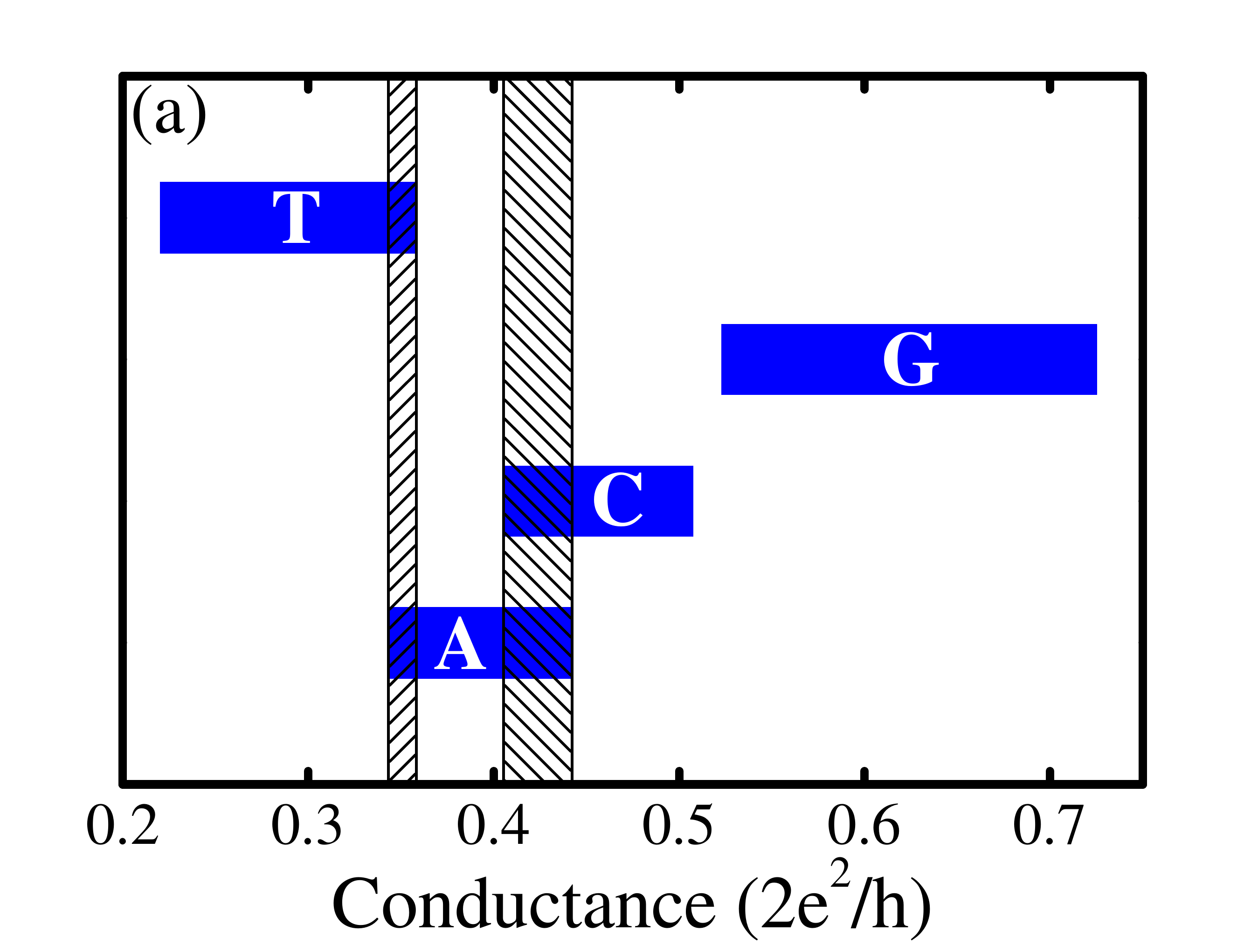}
\includegraphics[scale=0.37,angle=0]{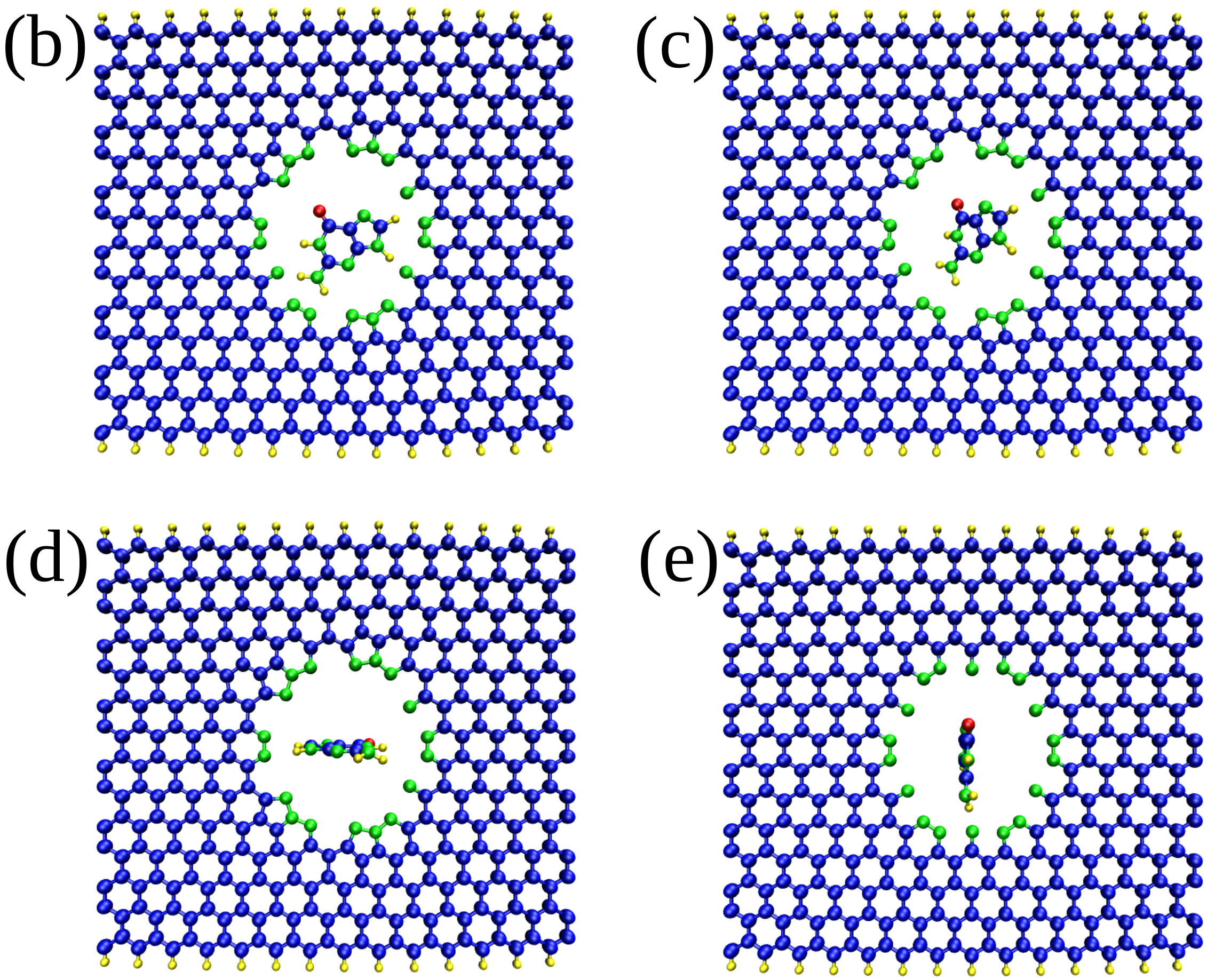}
\caption{(a) The variation of the room-temperature conductance of 14-ZGNR with N-pore due to the rotation of A, C, G, T nucleotides within the nanopore. The shaded vertical rectangles mark the regions of overlap between the conductance intervals associated with different DNA nucleotides.  (b)--(e) The specific positions of a nucleotide (guanine in the example) within the N-pore that define the conductance intervals shown in panel (a). The conductances in panel (a) were computed using our home-grown MT-NEGF-DFT code~\cite{Saha2010,Saha2009a}.}
\label{fig:fig5}
\end{figure}

Here we propose a novel device concept that could resolve these issues by  {\em abandoning the usage of small tunneling current altogether}. Its operation crucially relies on the existence of nanowires in which the spatial current profile~\cite{Zarbo2007} is confined around their transverse edges, so that drilling a nanopore should not change significantly their conductance which is of the order of few conductance quanta $2e^2/h$. When one of the four DNA nucleotides---adenine (A), cytosine (C), guanine (G), or thiamine (T)---is inserted into the nanopore in the course of DNA translocation, it will affect the charge density around the pore via its electrostatic potential thereby modulating {\em edge conduction currents} that are several orders of magnitude larger than tunneling currents across nanogaps~\cite{Zwolak2008,Tsutsui2010,Huang2010,Prasongkit2011,He2011,Postma2010} or nanopores~\cite{Nelson2010}.

The candidate nanowires supporting edge currents can be found among GNRs with zigzag edges or the very recently fabricated~\cite{Tao2011}  chiral GNRs, as well as among two-dimensional topological insulators (2D TI)~\cite{Hasan2010}. In the case of zigzag or chiral GNRs,  spatial profile of local currents carried by electrons around the charge neutral point (CNP) shows large magnitude around the edge~\cite{Areshkin2007a} and a tiny current flowing through their  interior. In 2D TI nanowires, similar situation will appear if the wire is narrow enough so that helical edge states overlap slightly and edge currents can be modulated. Otherwise, in sufficiently wide 2D TI wires current is strictly confined to the edges and cannot be affected by time-reversal-preserving impurities, vacancies or modulation of charge density because of the fact that helical edge states guide electrons of opposite spin in opposite directions to prevent their backscattering~\cite{Hasan2010}.

The recent proliferation of nanofabrication techniques~\cite{Tao2011,Cai2010,Jia2009} for GNRs with ultrasmooth edges are making them widely available, and their exposed surface allows for an easy integration into biosensors. Therefore, in the device depicted in Fig.~\ref{fig:setup} we choose GNR with zigzag edges~\cite{Jia2009}. The device corroborates the general modulation-of-edge-currents concept discussed above, as demonstrated by our central result in Fig.~\ref{fig:fig2} obtained via first-principles quantum transport simulations using two completely different~\cite{Saha2010,Saha2009a,quantumwise} computational implementations of the nonequilibrium Green function coupled to density functional theory (NEGF-DFT) formalism~\cite{Taylor2001,Areshkin2010}. 

Figure~\ref{fig:fig2} shows how each DNA nucleotide inserted in the center of the nanopore within the $yz$-plane [Fig.~\ref{fig:fig5}(e)] will change the device room-temperature conductance by a specific amount. When spatial orientation of nucleotides with respect to the pore is changed as in Fig.~\ref{fig:fig5}(b)--(d), the conductance will vary within the intervals shown in Fig.~\ref{fig:fig5}(a). The one-atom-thick GNRs make it possible to evade situations where several nucleotides inside the nanopore affect transverse conduction simultaneously, as could be the case when membrane carrying the nanopore is insufficiently thin~\cite{He2010}. The DNA nucleotide-specific modulation of current $I$ is achieved while remaining in the linear-response regime, where $I=GV$ is of the order of $\mu$A at bias voltage \mbox{$\simeq 0.1$ V}. Such sizable operating current is expected to be much larger than electrical noise due to thermal fluctuations of the DNA structure.

In the NEGF-DFT formalism~\cite{Taylor2001,Areshkin2010}, the Hamiltonian is not known in advance and has to be computed by finding converged spatial profile of charge via the self-consistent DFT loop for the density matrix \mbox{${\bm \rho} = \frac{1}{2 \pi i} \int dE\, {\bf G}^<(E)$} whose diagonal elements give charge density~\cite{Areshkin2010}. The NEGF formalism for steady-state transport operates with two central quantities, retarded ${\bf G}(E)$ and lesser Green functions ${\bf G}^<(E)$, which describe the density of available quantum states and how electrons occupy those states, respectively. In the coherent transport regime (i.e., in the absence of electron-phonon or electron-electron dephasing processes), only the retarded Green function is required to post-process the result of the DFT loop  by expressing the zero-bias electron transmission function between the left (L) and the right (R) electrodes as:
\begin{equation}\label{eq:transmission}
\mathcal{T}(E) = {\rm Tr} \left\{ {\bm \Gamma}_R (E)  {\bf G}(E) {\bm \Gamma}_L (E)  {\bf G}^\dagger(E)  \right\}.
\end{equation}
The matrices \mbox{${\bm \Gamma}_{L,R}(E)=i[{\bm \Sigma}_{L,R}(E) - {\bm \Sigma}_{L,R}^\dagger(E)]$} account for the level broadening due to the coupling to the electrodes, where ${\bm \Sigma}_{L,R}(E)$ are the self-energies introduced by the ZGNR electrodes~\cite{Areshkin2010}. The retarded Green function matrix of the central region is given by \mbox{${\bf G}=[E{\bf S} - {\bf H} - {\bm \Sigma}_L - {\bm \Sigma}_R]^{-1}$}, where in the local orbital basis $\{ \phi_i \}$ Hamiltonian matrix ${\bf H}$ is composed of elements \mbox{$H_{ij} = \langle \phi_i |\hat{H}_{\rm KS}| \phi_{j} \rangle$} and $\hat{H}_{\rm KS}$ is the effective Kohn-Sham Hamiltonian obtained from the DFT self-consistent loop. The overlap matrix ${\bf S}$ has elements \mbox{$S_{ij} = \langle \phi_i | \phi_j \rangle$}.

The conductance at finite temperature $T$ is obtained from the transmission function $\mathcal{T}(E)$ using the standard Landauer formula for two-terminal devices
\begin{equation}\label{eq:conductance}
G=\int\limits_{-\infty}^{+\infty} dE\, {\mathcal T}(E) \left(-\frac{\partial f}{\partial E}\right),
\end{equation}
where $f(E)=\{ 1 + \exp[(E-\mu)/k_BT] \}^{-1}$ is the Fermi function of the macroscopic reservoirs into which semi-infinite ideal leads terminate. The electrochemical potential $\mu$ is the same for both reservoirs at vanishingly small bias voltage.

The retarded Green function ${\bf G}$ is computed for the central region finite-ZGNR+nanopore of the biosensor shown in Fig.~\ref{fig:setup} consisting of around 700 atoms. This central region is attached to two semi-infinite ZGNRs electrodes of the same width. Whereas graphene is mechanically strong, it can be used  as both the membrane material carrying a nanopore and the electrode material. In real devices, ZGNR electrodes will eventually need to be connected to metallic electrodes attached to an external battery. However, the fact that GNRs used in experiments are typically rather long and screening takes place over a distance much shorter~\cite{Areshkin2010} than the central region  justifies the usage of semi-infinite ZGNRs as two electrodes in our simulations.

The edge carbon atoms will catch any bond partner they can possibly get to saturate their dangling bonds. We assume that ZGNR edges are passivated by hydrogen, while edge atoms of the nanopore can be bonded covalently to either hydrogen (H-pore) or nitrogen (N-pore). Prior to transport calculations, we use DFT to relax the coordinates of all atoms within finite-ZGNR+nanopore or finite-ZGNR+nanopore+nucleotide  until the forces on individual atoms are minimized to be smaller than $0.05$ eV/\AA$^2$. The converged result of this procedure is illustrated in Fig.~\ref{fig:fig5}(b)--(d) which shows how carbon and hydrogen atoms around the nanopore move away from it so that the edge of ZGNR acquires a slight curvature.

\begin{figure}
\includegraphics[scale=0.3,angle=0]{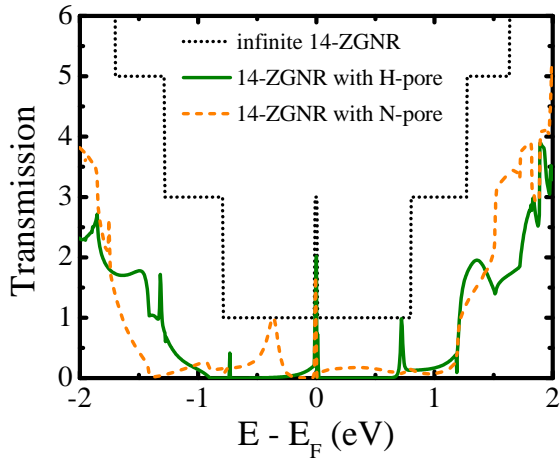}
\caption{The zero-bias electronic transmission function Eq.~(\ref{eq:transmission}) for an infinite homogeneous 14-ZGNR, whose edge carbon atoms are passivated by hydrogen, and the same nanoribbon with empty H-pore or N-pore of diameter $\approx 1.2$ nm (see Fig.~\ref{fig:setup}) drilled in its interior.}
\label{fig:fig3}
\end{figure}

The early theoretical studies of ZGNR-based devices have utilized a simplistic tight-binding model~\cite{Rycerz2007}  with single $\pi$ orbital per site and nearest neighbor hopping only, or its long-wavelength (continuum) approximation---the  Dirac-Weyl Hamiltonian~\cite{Brey2006}---valid close to CNP. However, to make connection to realistic device applications requires to take into account charge transfer~\cite{Areshkin2010} between different atoms~\cite{Cervantes-Sodi2008} that can be used to passivate edges or chemically functionalize graphene, as well as the charge redistribution~\cite{Areshkin2010} when finite bias voltage is applied. For example, the tight-binding model with the nearest-neighbor hopping predicts~\cite{Zarbo2007,Rycerz2007} that zero-temperature conductance of an infinite homogeneous ZGNR is $G=2e^2/h$ around the CNP and that current density profile is peaked in the middle of ZGNR despite transverse part of the eigenfunctions having maximum around the edges~\cite{Zarbo2007}.

\begin{figure}
\includegraphics[scale=0.35,angle=0]{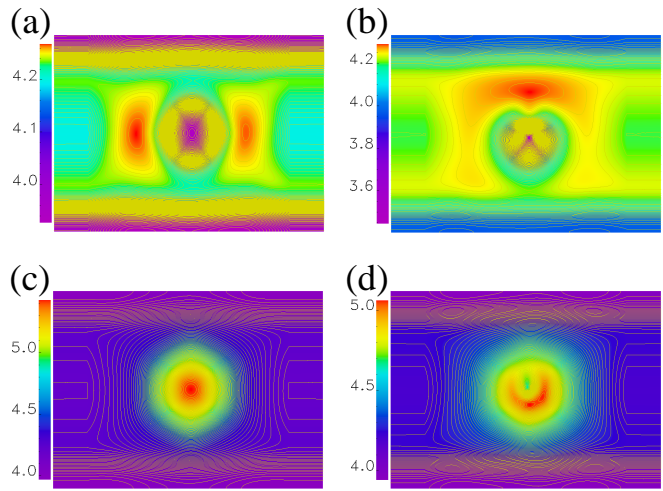}
\caption{The self-consistent Hartree potential at zero bias voltage for the central region of 14-ZGNR biosensors (Fig.~\ref{fig:setup}) with: (a) empty H-pore; (b) H-pore with cytosine positioned in its center within the $yz$-plane [Fig.~\ref{fig:fig5}(e)]; (c) empty N-pore; and (d) N-pore with thymine positioned in its center within the $yz$-plane [Fig.~\ref{fig:fig5}(e)].}
\label{fig:fig4}
\end{figure}

On the other hand, first-principles methods find that the zero-temperature conductance of an infinite homogeneous ZGNR is $G=6e^2/h$ around the CNP while local current is confined to flow mostly around the zigzag edges~\cite{Areshkin2007a}. This is illustrated by quantized steps in the transmission function in Fig.~\ref{fig:fig3} where ${\mathcal T}=3$ around the Fermi energy $E-E_F=0$, and the zero-temperature conductance is given by the simplified version of Eq.~(\ref{eq:conductance}), $G=\frac{2e^2}{h}{\mathcal T}(E)$. In the absence of any DNA base, the transmission function $\mathcal{T}(E)$ plotted in Fig.~\ref{fig:fig3}  remains large \mbox{$\mathcal{T} \simeq 2$} around CNP $E-E_F=0$ for an infinite ZGNR with either H-pore or  N-pore. This finding corroborates our conjecture that nanopore in the interior of a ZGNR is not able to substantially modify the current flow inherited from a  homogeneous nanoribbon since the local current density is mostly confined around the edges for electrons injected at energies sufficiently close to $E-E_F=0$. We note that using spin-unrestricted DFT reveals the presence of edge magnetic ordering and the corresponding band gap opening in ZGNRs which, however, is easily destroyed at room temperature~\cite{Yazyev2008,Kunstmann2011} so that for realistic device operation ZGNRs can be considered to be metallic~\cite{Kunstmann2011}.

The change in the room-temperature conductance of empty nanopores in Fig.~\ref{fig:fig2}(a) and nanopores with inserted nucleotide in Fig.~\ref{fig:fig2}(b) is more pronounced when the pore is terminated with nitrogen. Since reliability of predictions of NEGF-DFT simulations requires careful selection of the basis set and pseudopotentials in the DFT part of the calculation~\cite{Strange2008}, Fig.~\ref{fig:fig2} plots conductances obtained using two different computational implementations of the NEGF-DFT formalism. Our home-grown MT-NEGF-DFT code~\cite{Saha2010,Saha2009a} utilizes  ultrasoft pseudopotentials and Perdew-Burke-Ernzerhof (PBE) parametrization of the generalized gradient approximation (GGA) for exchange-correlation functional of DFT. The localized basis set is constructed from atom-centered orbitals (six per C atom, four per H atom, 8 per N atom, and 8 per O atom)  that are optimized variationally for the electrodes and the central molecule separately while their electronic structure is obtained concurrently. For comparison, we also used commercial ATK code~\cite{quantumwise} where pseudoatomic local orbitals are single-zeta polarized on C and H atoms and double-zeta polarized on N and O atoms.  In the case of ATK, we use Troullier-Martins norm-conserving  pseudopotentials and Perdew-Zunger (PZ) parametrization of the local density approximation (LDA) for the exchange-correlation functional of DFT. Importantly, both first-principles simulations yield very similar results for the conductance, as demonstrated in Fig.~\ref{fig:fig2}.

To explain the mechanisms by which DNA bases modulate charge transport in a ZGNR with a nanopore, we plot in Fig.~\ref{fig:fig4} the self-consistent Hartree potential within the central region of our biosensor at zero bias voltage obtained by solving the Poisson equation with the boundary conditions that match the electrostatic potentials of two attached ZGNR electrodes. We see that there is a substantial  difference in this potential when switching from an empty pore to nanopore containing a DNA nucleotide. In the examples in Fig.~\ref{fig:fig4}, cytosine is inserted into the H-pore and thymine into the N-pore---these are the situations for which there is the largest change in conductance in Fig.~\ref{fig:fig2} when compared to the corresponding empty nanopores.

An important issue~\cite{Prasongkit2011,Nelson2010} for the uniqueness of the conductance modulation signal associated with each nucleotide is to examine how such signal gets modified when varying the orientations of the nucleotide within a nanopore. For selected orientations shown in Fig.~\ref{fig:fig5}(b)--(e), the conductance variation for all four DNA nucleotides is plotted in Fig.~\ref{fig:fig5}(a). We find small overlap between conductance distribution for T and A or A and C, and no overlap between conductance intervals for T and C or C and G. Nevertheless, the intervals in Fig.~\ref{fig:fig5}(a) should be considered as 
setting only the limits on conductance variation since not all values within the interval will be sampled experimentally. 
That is, some of the nucleotide positions in Fig.~\ref{fig:fig5}(b)--(e) are selected to generate maximum conductance variation, and they would require significant bending of the DNA molecule to put the nucleotide into such position with respect to the nanopore. Instead, to find the most probable  fluctuations in orientations of DNA nucleotides,  the NEGF-DFT calculations of conductance should be coupled to molecular dynamics simulations of DNA translocation through the nanopore including hydrodynamic interactions with the
surrounding solvent~\cite{Fyta2011}.

\begin{figure}
\includegraphics[scale=0.3,angle=0]{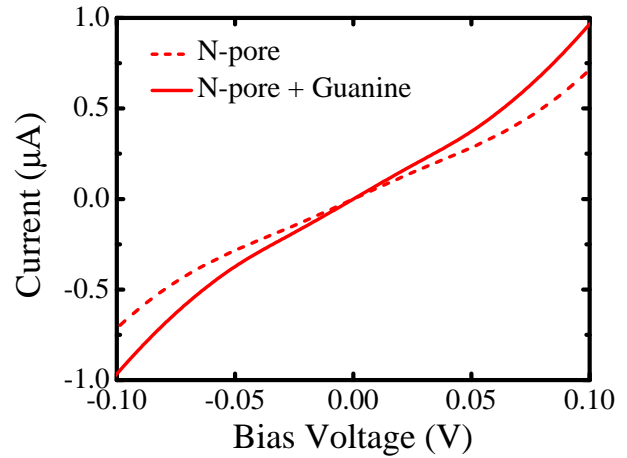}
\caption{Current-voltage characteristics of 14-ZGNR with N-pore which is empty (dashed line) or contains guanine (solid line) in its center placed within the $yz$-plane [Fig.~\ref{fig:fig5}(e)]. The current at finite bias voltage is computed using our home-grown MT-NEGF-DFT code~\cite{Saha2010,Saha2009a}.}
\label{fig:fig6}
\end{figure}

Finally, in Fig.~\ref{fig:fig6} we clarify the range of operating bias voltages that ensures a linear-response regime for our biosensor where the measured current is given simply by multiplying conductances in Figs.~\ref{fig:fig2} and ~\ref{fig:fig5} by
the bias voltage. Both current-voltage characteristics in Fig.~\ref{fig:fig6}, computed for a biosensor with an empty N-pore and the same pore containing guanine, behave linearly within the interval \mbox{$\simeq -0.05$ V} to \mbox{$\simeq 0.05$ V}.

In conclusion, using first-principles quantum transport simulations, we investigated a novel type of graphene nanopore-based sensors for rapid DNA sequencing which rely on nucleotide-specific modulation of a large transverse conduction current (of the order of $\mu$A at bias voltage \mbox{$\simeq 0.1$ V}). This is achieved by exploiting unique features of the electronic transport through graphene nanoribbons with zigzag edges where local current density is confined mostly around the nanoribbon edges. Another candidate nanowire carrying edge currents are recently fabricated~\cite{Tao2011} chiral GNRs. Thus, the nanopore in the GNR interior cannot substantially diminish the edge currents, whose magnitude is then modulated by the passage of nucleotides in the course of DNA translocation through the pore. Our analysis demonstrates that  each DNA nucleotide will generate  a unique electrostatic potential that modulates the charge density in the surrounding area. The operating current, which is several orders of magnitude greater than the tunneling current employed in previously considered biosensors with transverse electron transport~\cite{Postma2010,Prasongkit2011,He2011,Nelson2010,Zwolak2008,Meunier2008,Tsutsui2010,Huang2010} is expected to be much larger than its fluctuations due to thermal vibrations of the graphene membrane or structural fluctuations of the translocated DNA molecule. Furthermore, the device remains in the linear-response regime for bias voltages $\lesssim 0.05$ V.

\begin{acknowledgments}
We thank S. Sanvito for illuminating discussions. This work was supported by DOE Grant No. DE-FG02-07ER46374 (K. K. S. and B. K. N.) and NIH Grant No. R21HG004767 (M. D.). The supercomputing time was provided in part by the NSF through TeraGrid resource TACC Ranger under Grant No. TG-DMR100002.
\end{acknowledgments}





\end{document}